# Human cell recovery after microwave irradiation

Y.G. Shckorbatov, V.N. Pasiuga, V.A. Grabina, N.N. Kolchigin, D.D. Ivanchenko, V.N.Bykov

*Kharkiv National University, pl. Svobody, 4, Kharkiv, 61077, Ukraine,*

Cells of human buccal epithelium of 6 male donors were exposed to microwave radiation (frequency f=36,64 GHz, power density E=10, 100, and 400 $\mu W/cm^2$). Exposure time in all experiments was 10 seconds. Heterochromatin was stained by 2% orcein in 45 % acetic acid. The stainability of cells with trypan blue (0,5 %) and indigocarmine (5 mM) after 5 min of staining was investigated. Irradiation induced chromatin condensation (increase of number of heterochromatin granules) and increase of membrane permeability to vital dyes trypan blue and indigocarmine. Isolated human buccal cells had shown the ability to recover these changes. Number of heterochromatin granules lowered to initial level after 0,5 hour (E=10 $\mu W/cm^2$) and 2 hours (E=100, and 400 $\mu W/cm^2$) after irradiation. Cell plasma membrane permeability recovered a bit later, in correspondence, after 1 hour and 3 hours after irradiation.

## 1. Introduction

Phenomenon of recovery of cells isolated from organism was investigated in different aspects. The phenomenon of cell recovery after action of different external factors was thoroughly investigated by D. Nassonov and by his scientific school in at 1930-1960-th [1,2]. At present the problems of cell recovery are very intensively investigated in connection with mechanisms of DNA repair. Other mechanisms of cell repair are investigated less intensively. As an example of modern study in the area of cell biology of recovery may be investigation of effect of thymosin-ß4 treatment on human corneal epithelial cells exposed to ethanol in vitro. The efficacy of thymosin-ß4 in preventing mitochondrial disruption and in inhibiting caspasemediated apoptosis was examined. Nontransformed human corneal epithelial cells (HCECs) at passage 4 were untreated or treated with ethanol (20% for 20 seconds) or a combination of ethanol and thymosin-ß4. The cells were allowed to recover from ethanol treatment for 24 hours. Thymosin-ß4 treatment decreased deleterious mitochondrial alterations, significantly decreased cytochrome c release from mitochondria, and increased Bcl-2 expression in ethanol-exposed human corneal epithelial cells. In ethanol-exposed corneal epithelium thymosin-ß4 treatment inhibited caspase-2, -3, -8, and -9 activity, with caspase-8 showing the most significant inhibition. Thymosin-ß4 treatment resulted in no significant effect on the proliferation of human corneal epithelial cells after ethanol exposure (Sosne et al., 2004).

Detailed investigations of human cell membrane recovery after the influence of different bacterial toxins were realized in studies of scientific group from Johannes Gutenberg University, Mainz, Germany [4-6]. A moderate concentration (1-50 micrograms/ml) of staphylococcal alpha-toxin similarly produces a small pore in membranes of human fibroblasts, rapid leakage of K+ and a drop in cellular ATP to 10-20% of normal levels in 2 h. In the presence of medium supplemented with serum and at pH 7.4, the cells are able to recover from toxin attack, so that normal levels of K+ and ATP are reached after 6-8 h at 37 degrees C. The repair process is dependent on the presence of serum in the medium and is very sensitive to pH [4]. In the next work authors studied the influence on human epithelial cells of nanomolar concentrations of chromatographically purified *Serratia marcescens* hemolysin (ShlA) which caused vacuolation and subsequent lysis of the cells. ShlA-treated cells were depleted of ATP, and small pores were formed in cell membranes. ATP depletion in ShlA-treated cells was reversed by a change to a medium free of ShlA. This restoration of the initial ATP level depended on the time at which the medium was changed. After the change of medium, the ATP level decreased over the first 30 min and then increased continuously. This time course of ATP level restoration indicates that the fibroblasts and epithelial cells are able to inactivate the toxin pores by possibly closing the pores, by proteolytic degradation of ShlA, or by a complete shedding of the toxin from the membrane. The repair of ShlA pores was inhibited by cycloheximide. These data suggest that the repair of ShlA pores is dependent on protein synthesis [5]. In the subsequent work authors following the observation that cells are able to recover from membrane lesions incurred by Staphylococcus aureus alpha-toxin and streptolysin O (SLO), investigated the role of p38 in this process. The p38 phosphorylation occurred in response to attack by both toxins, commencing within minutes after toxin treatment and waning after several hours. While SLO reportedly activates p38 via ASK1 and ROS, was shown that this pathway does not play a major role for p38 induction in alpha-toxin-treated cells. Strikingly divergent effects of p38 blockade were noted depending on the toxin employed. In the case of alpha-toxin, inhibition of p38 within the time frame of its activation led to disruption of the recovery process and to cell death. In contrast, blockade of p38 in SLO permeabilized cells did not affect the capacity of the cells to replenish their ATP stores [6].

In the present paper we analyze the possibility of cell recovery after the influence of low-level electromagnetic radiation. Previously we demonstrated that microwaves induce an increase of quantity of heterochromatin granules in human cell nuclei [7,8]. We also demonstrated that low-level microwave radiation induce increase of cell membrane permeability to vital dyes [9]. The fact of cell recovery after electromagnetic irradiation was shown earlier [10,11] but in this paper we study the dependence of dynamics of cell recovery from irradiation of different intensity of for the first time.

## 2. Materials and methods

### 2.1. Human cells

Cells of buccal epithelium were obtained from the inner surface of donor's cheek by light scraping with a blunt sterile spatula. This operation is absolutely bloodless and painless. All the good-will donors were informed about the purposes of investigation. Our investigations are performed in accordance with the European Convention on Human Rights and Biomedicine (1997), Declarations and Recommendations of the First, the Second and the Third National (Ukrainian) Congresses of Bioethics (Kiev, Ukraine, 2001, 2004, 2007) and Ukrainian legislation.

The cells were placed in solution of the following composition: 3.03 mM phosphate buffer (pH=7.0) with addition of 2.89 mM calcium chloride (Reachem, Moscow, Russia) and use for further experiments. 25 µl of cell suspension containing several thousand of cells were placed on the glass slide and subjected to microwave irradiation. Immediately after the irradiation procedure cells were stained with 2 % orcein (Merck AG, Darmstadt, Germany) solution in 45 % acetic acid (Reachem, Moscow, Russia), indigocarmine (Merck AG, Darmstadt, Germany) and trypan blue (Reachem, Moscow, Russia). Donors of cells were of male sex, non-smokers. Donor A was of 18 years old, donor B – 20 years old, donor C – 24 years old, donor D – 35 years old, E – 53 years old, and donor F – 55 years old.

### 2.2. Irradiation procedure

As a source of electromagnetic radiation of frequency f=36,64 GHz we applied a semi-conductor device. Irradiation was conducted at room temperature. In all experiments irradiation power density at the surface of exposed object was E = 10, 100, and 400 µW/cm$^2$. Exposure time in all experiments was 10 s.

### 2.3. Chromatin state evaluation

The number of heterochromatin granules was estimated by the method described earlier [12]. The suspended cells (2 µl) were placed on the cover slide and irradiated. Cells were stained with 2 % orcein in 45% acetic acid after 30 sec, 30 min, 1 hour and 2 hours after irradiation. Orcein is specific stain for heterochromatin staining as it was shown in [13]. Cells were investigated at magnification x 600. Each cell sample contained several thousands of cells. In each variant of experiment heterochromatin granules quantity (HGQ) was estimated in 30 cell nuclei and the mean HGQ value and the standard error of this value were calculated. This number of cells (30) was determined in our previous experiments as an optimal for such analysis. The variability of HGQ in cell sample gives the value of the standard error of the mean data (SEM). In our experiments SEM was less than 5% of the mean HGQ. Every experiment was conducted in triplet for each donor, using three different cell samples obtained in different days.

### 2.4. Evaluation of the state of cell membrane

Indigocarmine (Merck AG, Darmstadt, Germany) and trypan blue (Reachem, Moscow, Russia) were used for cell damage assessment. These methods also may be considered as a method reflecting cell viability [14,15]. We have stained cells after 30 sec, 30 min, 1 hour, 2 hours and 3 hours after irradiation. We used the percentage of unstained cells after 5 min of staining with 5 mM indigo carmine and 0.5% trypan blue solutions in the buffer solution described above. In one experiment we analyzed 5 cell samples irradiated independently (N1). In each cell sample we analyzed 100 cells (N2) and determined the percentage of unstained cells for each cell sample. After this we calculated the mean number of unstained cells for all 5 cell samples and the standard error of this value.

### 2.5. Statistical analysis

All experimental results were statistically processed using Student's t-test. The probability level assumed in this paper is P<0,05.

## 3. Results and discussion

In our experiments microwave irradiation induced a significant increase of heterochromatin granule quantity (HGQ) immediately after cell sample exposure. After a definite period of recovery HGQ gradually decreased to control level, and the time of full recovery depended on the intensity of irradiation (Fig 1-6).

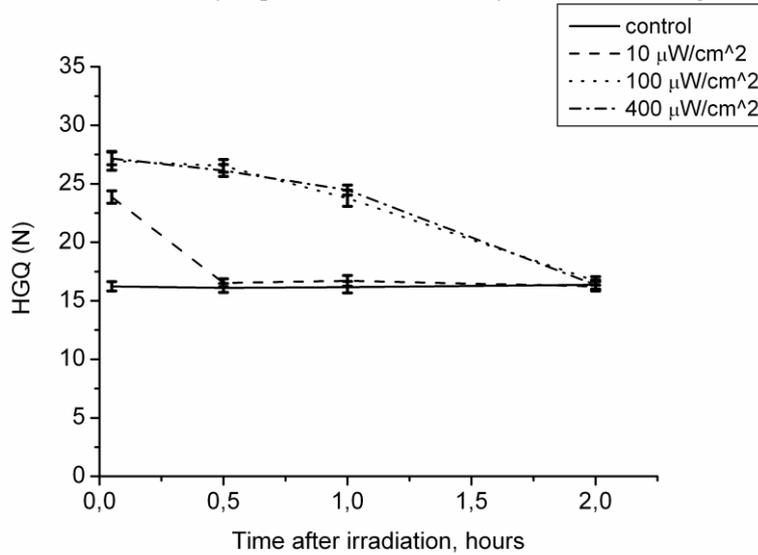

Fig. 1. Changes in heterochromatin granule quantity (HGQ) after microwave radiation in different periods after irradiation in cells of Donor A

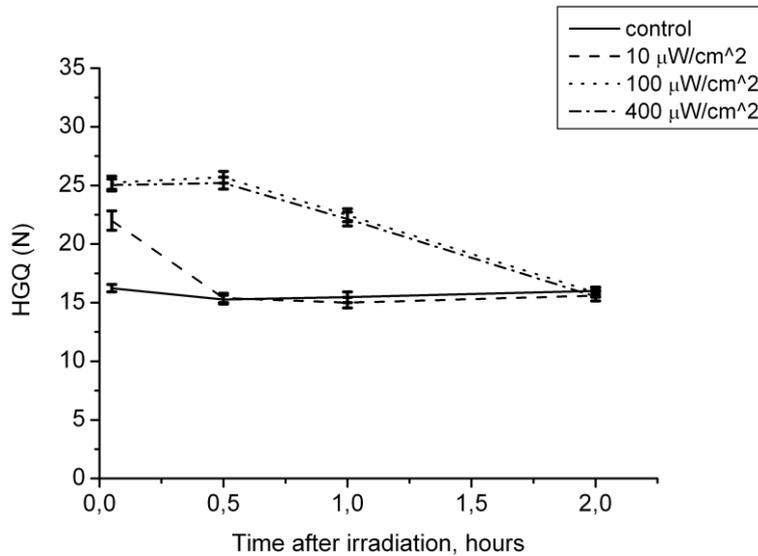

Fig. 2. Changes in heterochromatin granule quantity (HGQ) after microwave radiation in different periods after irradiation in cells of Donor B

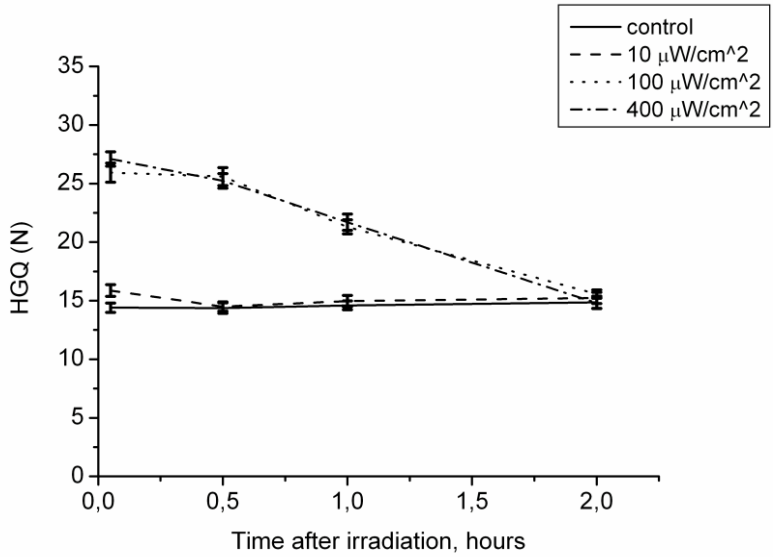

Fig. 3. Changes in heterochromatin granule quantity (HGQ) after microwave radiation in different periods after irradiation in cells of Donor C

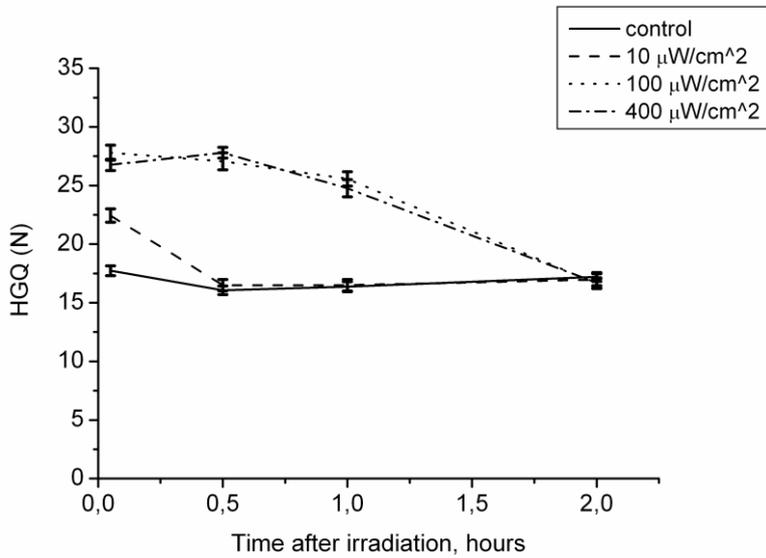

Fig. 4. Changes in heterochromatin granule quantity (HGQ) after microwave radiation in different periods after irradiation in cells of Donor D

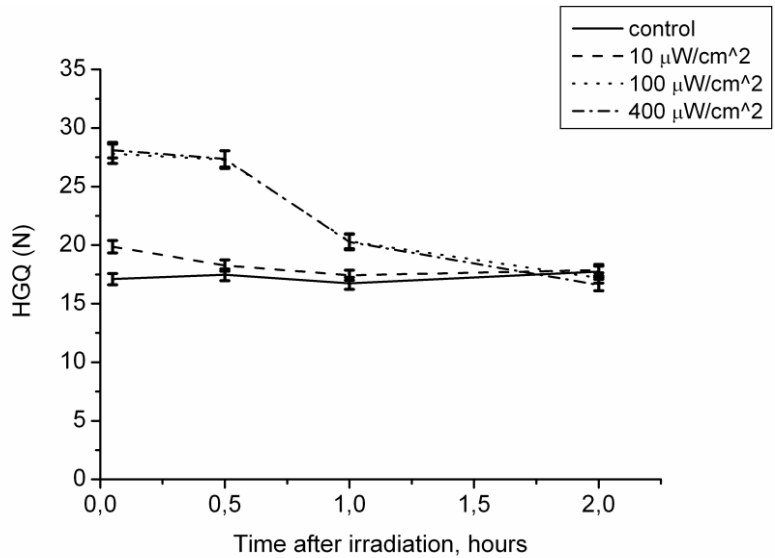

Fig. 5. Changes in heterochromatin granule quantity (HGQ) after microwave radiation in different periods after irradiation in cells of Donor E

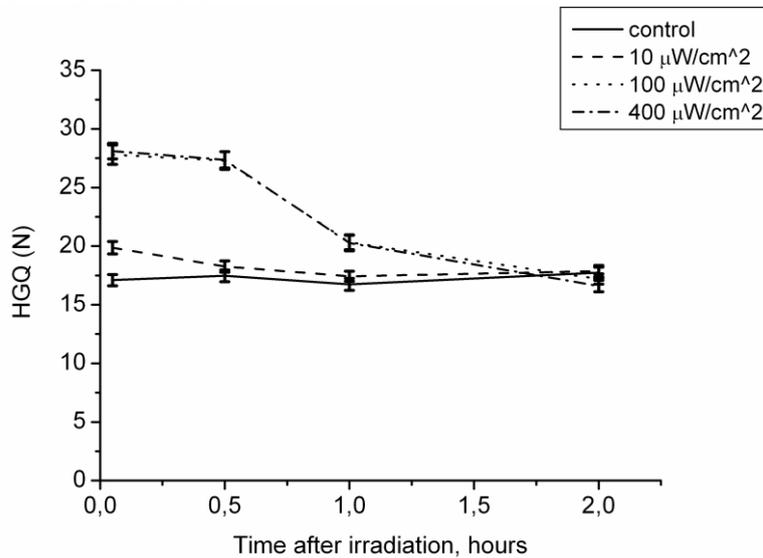

Fig. 6. Changes in heterochromatin granule quantity (HGQ) after microwave radiation in different periods after irradiation in cells of Donor F

Our experimental data show an increase of heterochromatin granule quantity after cell sample exposure to low-level microwave irradiation. For understanding the biological meaning of the effect of chromatin condensation in granules some additional explanations may be done.

Chromatin is DNA-protein complex in the interphase nucleus forming chromosomes in the phases of cell division. The main parts of chromatin are euchromatin and heterochromatin. Euchromatin presents most transcriptional active part of chromatin and heterochromatin is transcriptional repressed part of chromatin. Translocation of euchromatin to heterochromatin and vice versa is a structural representation of changes in chromatin activity, condensation of chromatin, so-called heterochromatinization, is connected with the process of decrease of the functional activity of chromatin [16]. In our previous works we demonstrate induced by electromagnetic field condensation of chromatin in granules [7,8,10] but the process of recovery of microwave induced changes in chromatin microscopic morphology remained almost uninvestigated.

As one can see from Fig 1-6, in general the reaction of cells of all donors to irradiation was the same. Immediately after irradiation HGQ increased and this increase was more pronounced after cell exposure to irradiation of intensity 100 and 400 $\mu W/cm^2$, than 10 $\mu W/cm^2$. After a period of recovery HGQ decreased to control level. Time of recovery was 0,5 hour for intensity 10 $\mu W/cm^2$ and 2 hours if intensity of irradiation was 100 and 400 $\mu W/cm^2$. Cells of donor C proved to be more stable to the irradiation of intensity 10 $\mu W/cm^2$ – the HGQ in these cells remained almost unchanged.

In the next series of experiments we studied changes of cell membranes to dye trypan blue (Fig. 7-12). Trypan blue is commonly used in cell biology to access viability of cells because normal living cell do not absorb this dye and it penetrates only in cells with damaged plasma membrane [15].

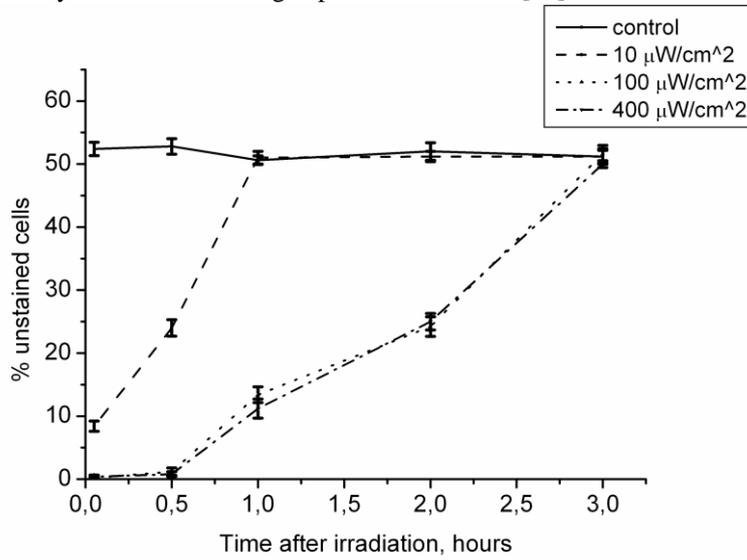

Fig. 7. Changes in cell membrane permeability for trypan blue after microwave radiation exposure in different periods after irradiation (Donor A)

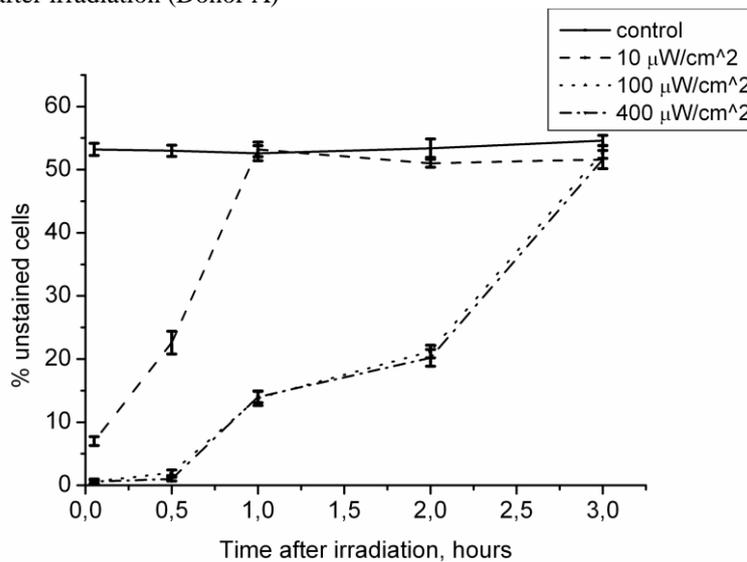

Fig. 8. Changes in cell membrane permeability for trypan blue after microwave radiation exposure in different periods after irradiation (Donor B)

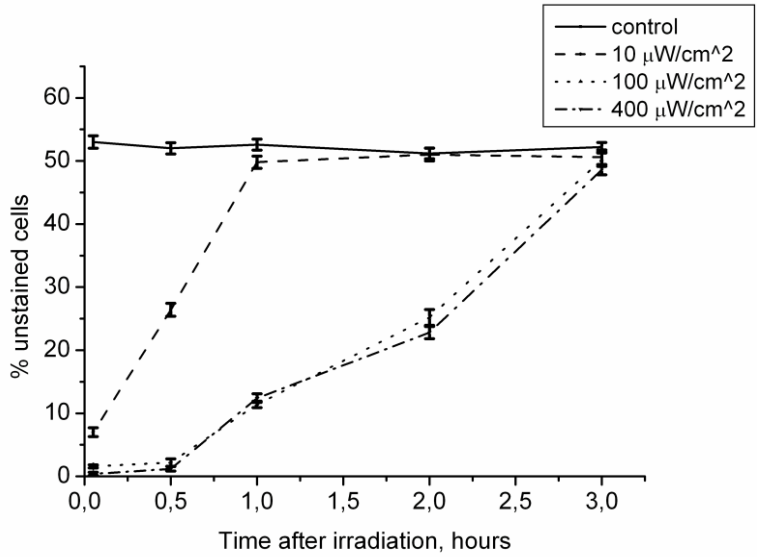

Fig. 9. Changes in cell membrane permeability for trypan blue after microwave radiation exposure in different periods after irradiation (Donor C)

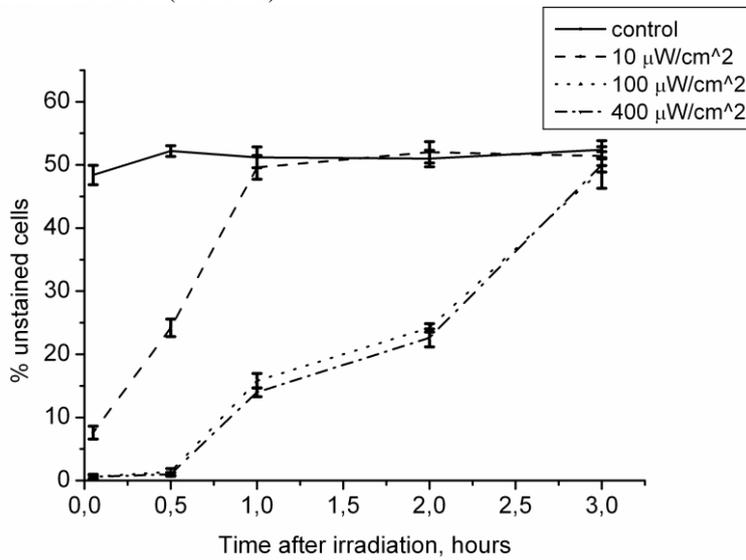

Fig. 10. Changes in cell membrane permeability for trypan blue after microwave radiation exposure in different periods after irradiation (Donor D)

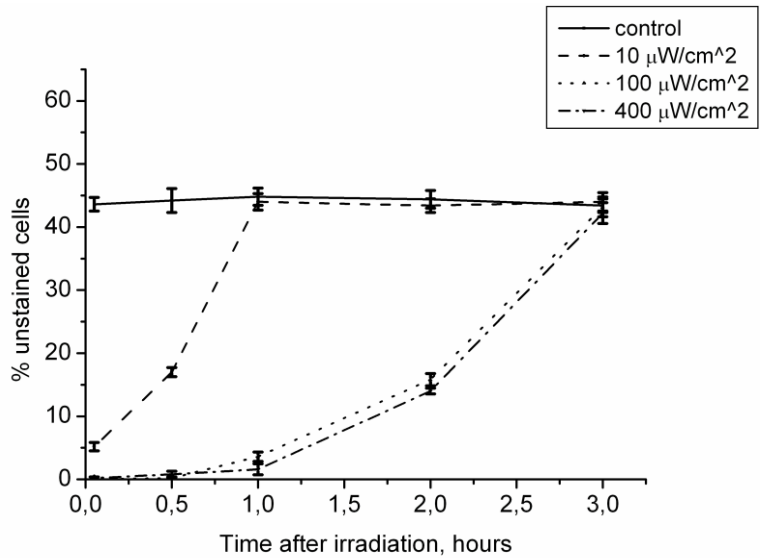

Fig. 11. Changes in cell membrane permeability for trypan blue after microwave radiation exposure in different periods after irradiation (Donor E)

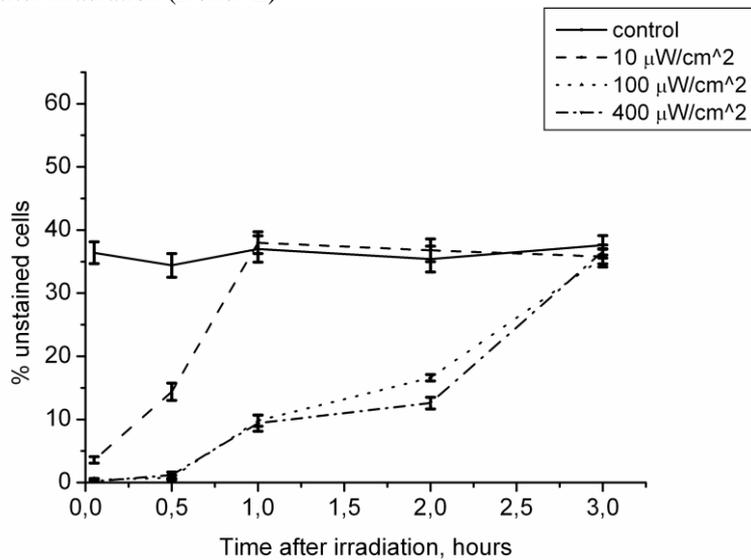

Fig. 12. Changes in cell membrane permeability for trypan blue after microwave radiation exposure in different periods after irradiation (Donor F)

In our experiments microwave irradiation decreased the percentage of unstained cells, and the effect of irradiation depended on its intensity.

The data presented in Fig. 7-12 indicate that in control level of trypan blue staining differed in cells of different donors: in young donors (A-C) it was more than in older donors (D-F). This observation agrees with our previous data indicating increase of cell membrane permeability with age [14]. The process of cell recovery from irradiation of intensity 10 $\mu W/cm^2$ takes 1 hour, and for irradiation of intensity 100 and 400 $\mu W/cm^2$ it takes 3 hours.

In purpose to determine general regularities in the process of cell membrane recovery after microwave irradiation we also studied cell membrane permeability changes to vital dye indigocarmine (Fig. 13-18). In low concentrations this dye also penetrates in damaged cells, undamaged cells remain unstained as it demonstrated in (Shckorbatov et al., 1995).

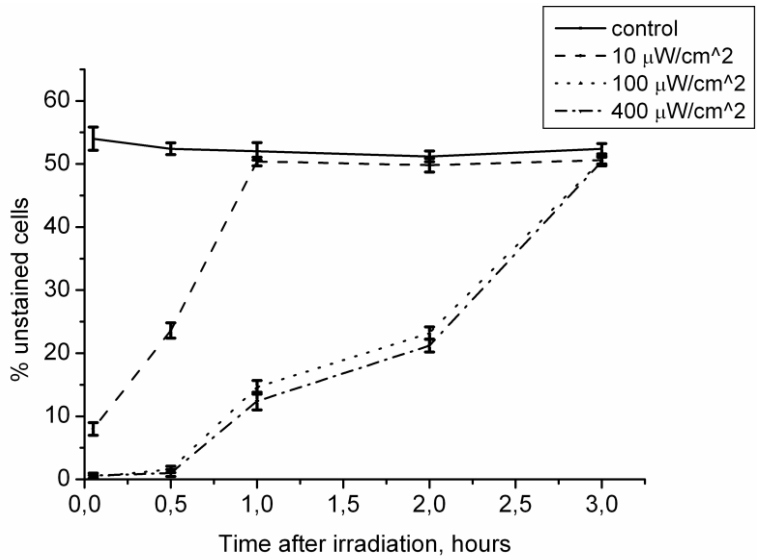

Fig. 13. Changes in cell membrane permeability for indigocarmine after a microwave radiation exposure in different periods after iirradiation (Donor A)

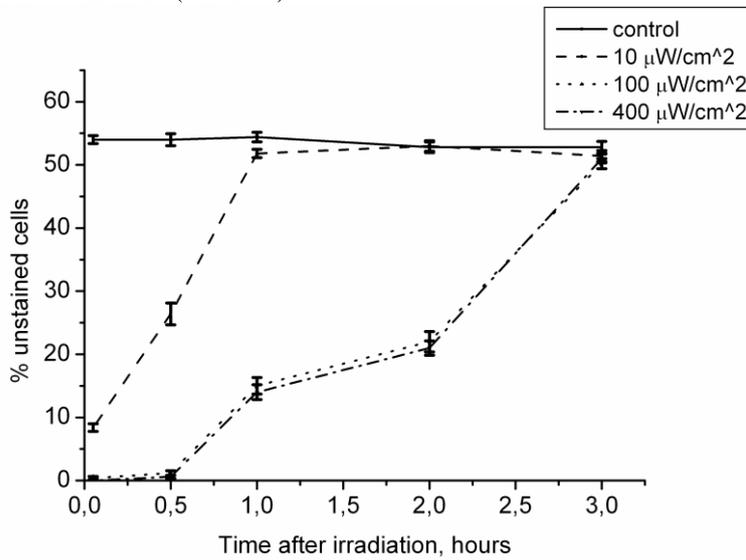

Fig. 14. Changes in cell membrane permeability for indigocarmine after a microwave radiation exposure in different periods after irradiation (Donor B)

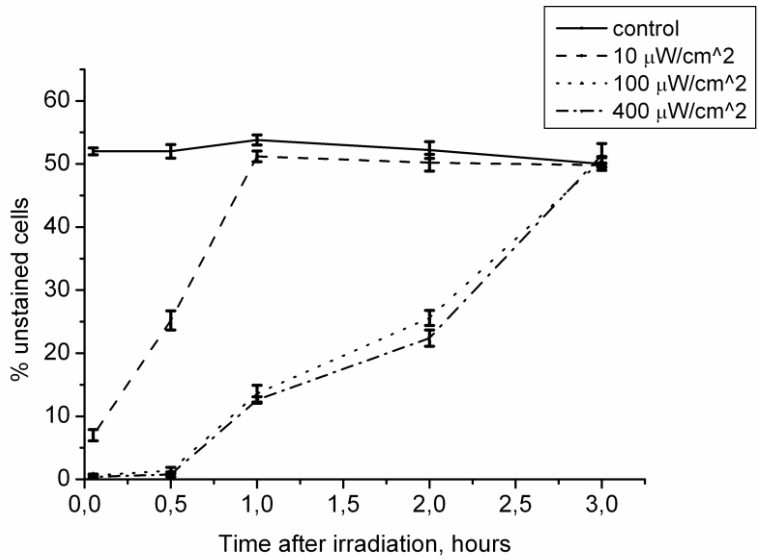

Fig. 15. Changes in cell membrane permeability for indigocarmine after a microwave radiation exposure in different periods after irradiation (Donor C)

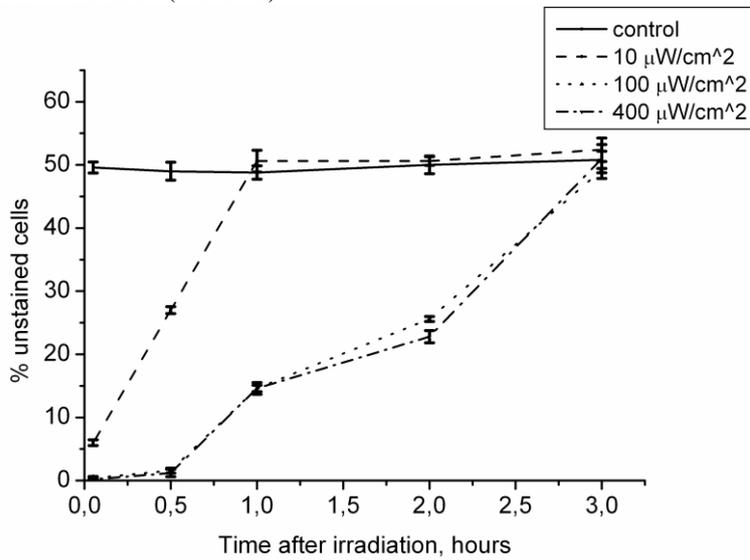

Fig. 16. Changes in cell membrane permeability for indigocarmine after a microwave radiation exposure in different periods after irradiation (Donor D)

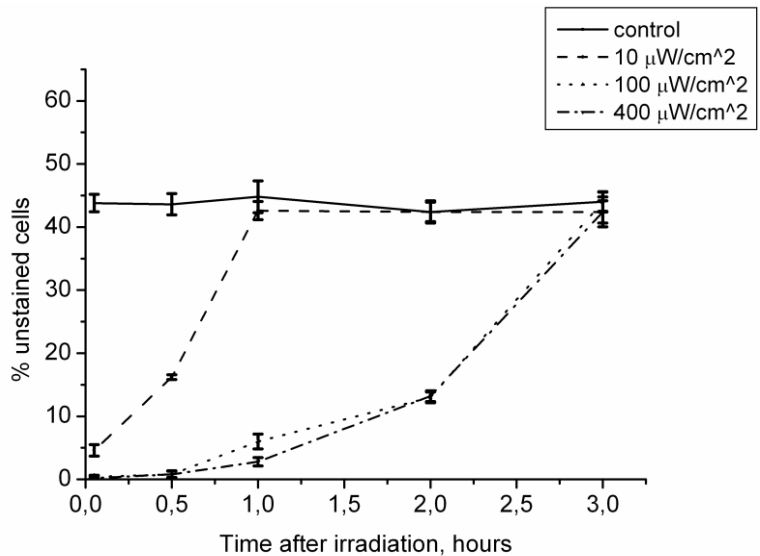

Fig. 17. Changes in cell membrane permeability for indigocarmine after a microwave radiation exposure in different periods after irradiation (Donor E)

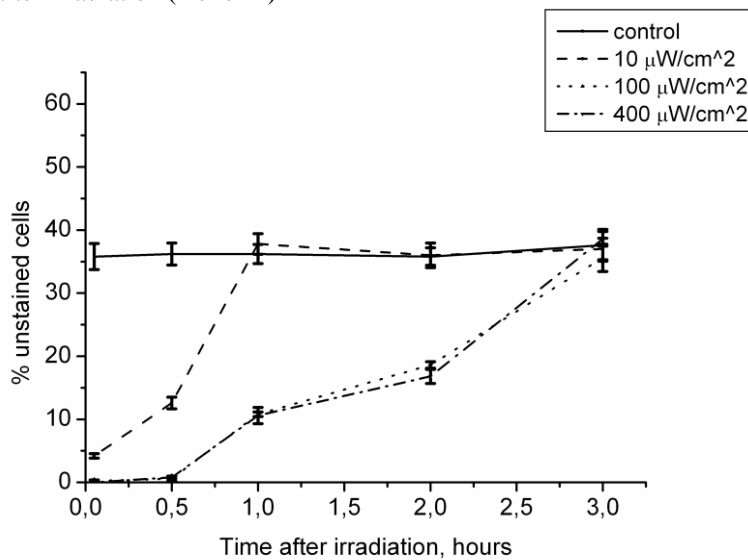

Fig. 18. Changes in cell membrane permeability for indigocarmine after a microwave radiation exposure in different periods after irradiation (Donor F)

As one can see reaction of cells stained with indigocarmine was almost similar to it in trypan blue stained cells. Cells of elder donors had a lower initial level of cell staining, irradiation induced decrease of percentage of unstained cells in cell samples of all donors. After a period of recovery (1 hour for intensity of irradiation 10 $\mu W/cm^2$ and 3 hours for intensity 100 and 400 $\mu W/cm^2$) the percentage of unstained cells became close to control level.

In our opinion the possibility of cell membrane recover after microwave radiation is connected to the general ability of cell to restore damages produced by external factors. The molecular bases of cell membrane recovery in our experiments remain uninvestigated. We speculate that this process may be connected with process of rearrangement of lipid molecules in outer cell membrane. But for the revealing of the concrete mechanism of this processes needs further investigations.

Comparing of the process of recovery of microscopic structure of chromatin and state of membrane integrity we can conclude that process of chromatin structure renewal is faster than membrane structure renewal. After irradiation at intensity 10 $\mu W/cm^2$ chromatin structure recover after 0,5 hour and initial membrane integrity after 1 hour. After irradiation at intensity 100 and 400 $\mu W/cm^2$ chromatin structure recover after 2 hours and initial membrane integrity after 3 hours. Such difference in times of recovery makes it possible to suppose the necessity of nuclear recovery for the process of membrane recovery. This is the theme of our further investigations.

## 4. Conclusions

Human cell exposure to low-level microwave radiation induced chromatin condensation (increase of number of heterochromatin granules) and increase of membrane permeability to vital dyes. These changes proved to be transitory and number of heterochromatin granules lowered to initial level after 0,5 hour and 2 hours, that depended on radiation intensity, and membrane permeability at the same irradiation doses recovered a bit later, after 1 hour and 3 hours.

## References

[1] Nassonov, D. N. (1930) Uber den Einfluss des Oxidationsprocesse auf die Verteilung von Vitalfarbs toffen in Zelle. Ztschr Zellforsch mikrosk Anat 11:179-217.
[2] Nassonov, D. N., Alexandrov, V. Ya. (1940). On the reaction of the living substance to the external influences. Moscow:Publishers of Academy of Sciences of USSR (in Russian).
[3] Sosne, G., Atif Siddiqi, A., Kurpakus-Wheater, M. (2004). Thymosin-ß4 Inhibits Corneal Epithelial Cell Apoptosis after Ethanol Exposure In Vitro. Investigative Ophthalmology & Visual Science 45(4):1095-1100.
[4] Walev, I., Palmer, M., Martin, E., et al. (1994). Recovery of human fibroblasts from attack by the pore-forming alpha-toxin of Staphylococcus aureus. Microb Pathog 17(3):187-201.
[5] Hertle, R., Hilger, M., Weingardt-Kocher, S., Walev, I. (1999). Cytotoxic Action of Serratia marcescens Hemolysin on Human Epithelial Cells. Infection and Immunity 67(2): 817-825.
[6] Husmann, M., Dersch, K., Bobkiewicz, W., et al. (2006). Differential role of p38 mitogen activated protein kinase for cellular recovery from attack by pore-forming S. aureus alpha-toxin or streptolysin O. Biochem Biophys Res 344(4):1128-1134.
[7] Shckorbatov, Y. G., Shakhbazov, V. G., Grigoryeva, N. N., Grabina, V. A. (1998). Microwave irradiation influences on the state of human cell nuclei. Bioelectromagnetics 19(7):414-419.
[8] Shckorbatov, Y. G., Pasiuga, V. N., Kolchigin, N. N., et al. (2009b). The influence of differently polarized microwave radiation on chromatin in human cells. International Journal of Radiation Biology 85(4):322-329.
[9] Shckorbatov, Y. G., Shakhbazov, V. G., Navrotskaya, V. V., et al. (2002). Electrokinetic properties of nuclei and membrane permeability in human buccal epithelium cells influenced by the low-level microwave radiation. Electrophoresis 23:2074-2079.
[10] Shckorbatov, Y. G., Pasiuga, V. N., Kolchigin, N. N., et al. (2009a). Changes in the human nuclear chromatin induced by ultra wideband pulse irradiation. Central European Journal of Biology 4(1):97-106.
[11] Pasiuga, V. N., Shckorbatov, Y. G., Kolchigin, N. N., et al. (2009).The process of recovery of cell membrane damage produced by the low-level microwave radiation. 7th International Conference on Antenna Theory and Techniques. Lviv:360-362.
[12] Shckorbatov, Y. G. (1999). He-Ne laser light induced changes in the state of chromatin in human cells. Naturwissenschaften 86(9):452-453.
[13] Sanderson A., Stewart J. Nuclear sexing with acetoorcein. Brit. Med. J., 1961, 2: 1065-1067.
[14] Shckorbatov, Y. G., Shakhbazov, V. G., Bogoslavsky, A. M., Rudenko, A. O. (1995). On age-related changes of cell membrane permeability in human buccal epithelium cells. Mech. Ageing Develop 83:87-90.
[15] Hodgson, E., ed. (2004). A textbook of modern toxicology. Third edition (pp.14-16). New Jersey: John Wiley & Sons, Inc.
[16] Lewin, B., (2004). Genes VIII. Ney York:Pearson Prentice Hall.
.